\documentclass[aps,pra,twocolumn,superscriptaddress,floatfix,longbibliography]{revtex4}
\usepackage{amsfonts}
\usepackage{amssymb}
\usepackage{graphicx}
\usepackage{dcolumn}
\usepackage{bm}
\usepackage{amsmath}
\usepackage[colorlinks,linkcolor=magenta,citecolor=blue,urlcolor=blue]{hyperref}
\usepackage{changes}
\begin{document}
	
	% Main title of the paper
	\title{Exploring Multifractal Critical Phases in Two-Dimensional Quasiperiodic Systems}
	\author{Chao Yang}
	\thanks{These authors contribute equally to this work.}
	\affiliation{Shenzhen Institute for Quantum Science and Engineering,
			Southern University of Science and Technology, Shenzhen 518055, China}
	\affiliation{International Quantum Academy, Shenzhen 518048, China}
	\affiliation{Guangdong Provincial Key Laboratory of Quantum Science and Engineering, Southern University of Science and Technology, Shenzhen 518055, China}
	
	\author{Weizhe Yang}
	\thanks{These authors contribute equally to this work.}
	\affiliation{Department of Physics, Southern University of Science and Technology, Shenzhen 518055, China}
	
	\author{Yongjian Wang}
	\thanks{Corresponding author: wangyongjian@amss.ac.cn}
	\affiliation{School of Mathematics and Statistics, Nanjing University of Science and Technology, Nanjing 210094, China}
	
	\author{Yucheng Wang}
	\thanks{Corresponding author: wangyc3@sustech.edu.cn}
	\affiliation{Shenzhen Institute for Quantum Science and Engineering,
		Southern University of Science and Technology, Shenzhen 518055, China}
	\affiliation{International Quantum Academy, Shenzhen 518048, China}
	\affiliation{Guangdong Provincial Key Laboratory of Quantum Science and Engineering, Southern University of Science and Technology, Shenzhen 518055, China}

\begin{abstract}
The multifractal critical phase (MCP) fundamentally differs from extended and localized phases, exhibiting delocalized distributions in both position and momentum spaces. The investigation on the MCP has largely focused on one-dimensional quasiperiodic systems. Here, we introduce a two-dimensional (2D) quasiperiodic model with a MCP. We present its phase diagram and investigate the characteristics of the 2D system's MCP in terms of wave packet diffusion and transport based on this model. We further investigate the movement of the phase boundary induced by the introduction of next-nearest-neighbor hopping by calculating the fidelity susceptibility. Finally, we consider how to realize our studied model in superconducting circuits. Our work opens the door to exploring MCP in 2D systems.
\end{abstract}

%\begin{keywords}
%	Anderson localization \sep critical phase \sep quantum simulation \sep transport \sep  
%\end{keywords}

\maketitle

\section{Introduction}
The interference of multiply scattered waves caused by disorder leads to the exponential decay of the wave function. This phenomenon is referred to as Anderson localization (AL)~\cite{Anderson1958,RMP1,RMP2,Kramer1993} and is widely present in disordered systems.  Additionally, quasiperiodic potentials can also induce AL, and have received widespread attention in both theoretical~\cite{Soukoulis1981,DasSarma1988,Biddle2009,XLi2017,HYao2019,Ganeshan2015,Wang1,XDeng2019,Ribeiro,YaruLiu} and experimental~\cite{Roati2008,Bloch4,An2018,JiasT,TXiao2021,Weld,HengFan} studies in recent years. Quasiperiodic systems exhibit physics distinct from randomly disordered systems.
Generally, in low-dimensional disordered systems, even a small
amount of disorder can cause AL. However, even in one-dimensional (1D) quasiperiodic systems,
Anderson transitions only occur when the quasiperiodic potential reaches a certain strength. This
leads to phenomena usually seen in three-dimensional disordered systems, such as mobility edges (MEs), appearing in 1D quasiperiodic systems~\cite{Soukoulis1981,DasSarma1988,Biddle2009,XLi2017,HYao2019,Ganeshan2015,Wang1,XDeng2019,Ribeiro,YaruLiu,Roati2008,Bloch4,An2018,JiasT}. The ME represents the critical energy that separates localized and extended states, describing the Anderson transition driven by changes in energy. The transition between localized and delocalized states can also occur across the entire spectrum~\cite{RPModel1,RPModel2}, meaning there is no ME, with the possibility of a critical point or even a multifractal critical phase (MCP). It was further discovered that in 1D quasiperiodic
systems, not only critical points exist, but MCP can also be present~\cite{TXiao2021,Weld,HengFan,WangYC2021,Kohmoto1990,WangMBC,Ribeiro2}.

MCP is a fundamental physical phase distinct from the extended phase and the localized phase. It exhibits various interesting features, such as special spectral statistics~\cite{energy1,energy2} and multifractal distribution of wave-functions~\cite{mul1,mul2}. 
From the perspective of wave packet dynamics (WPD), the WPD of extended phase and localized phase are ballistic and localized~\cite{Hiramoto,Geisel1997}, respectively. Critical phases, however, are more diverse. Critical phases induced by quasiperiodic potentials typically approach normal diffusion~\cite{Hiramoto,Geisel1997}, while critical phases in Fibonacci chains may also exhibit super-diffusion or sub-diffusion~\cite{FibonacciRMP}. 
From the perspective of transport, in the extended phase, the magnitude of conductance does not depend on the system size, while in the critical (localized) phase, conductance will decay in a power-law (exponential) form with increasing 1D system size~\cite{transport1,transport2,transport3,Wang2022}. it is interesting to note that multifractal critical states can also enhance the superconducting transition temperature~\cite{enhance1,enhance2,enhance3,enhance4} or the quantum metric~\cite{metric1}. 

Previous studies on MEs and MCPs in quasiperiodic systems mainly focused on
one dimension. In recent years, 2D quasiperiodic systems have gradually attracted attention~\cite{2D1,2D2,2D3,2D4,2D5,WangZhang,2DME0,2DME1,2DME2}. For example, research on MEs has been extended to 2D quasiperiodic systems~\cite{WangZhang,2DME0,2DME1,2DME2}.
However, studies on MCPs induced by quasiperiodic potentials in 2D systems have yet to emerge~\cite{noteadd}, leaving many of the properties of these systems unclear, such as how to perform finite-size scaling analysis on the eigenfunctions to determine the phase transition points and the associated critical exponents, and how the WPD and transport behaviors of the 2D MCP differ from those in the extended, localized, and one-dimensional cases.
%In this work, we extend the MCP to two-dimensional (2D) systems and investigate its dynamical diffusion and transport properties

In this work, we study a 2D quasiperiodic model. When the next-nearest-neighbor (NNN) hopping strength of this model is set to zero, this model can be separated into two 1D models with exact critical phases. Thus, we can determine the phase diagram of this model in the special case where the NNN hopping strength is zero. Using this special case as a benchmark, 
we explore a finite-size analysis
method and find that it can accurately determine the phase boundary locations, which are consistent with the phase boundaries obtained through decomposition. This demonstrates the feasibility of this finite-size analysis method in 2D quasi-periodic systems. We further investigate the WPD and transport behavior of critical phases in 2D systems, comparing them with the 1D case. Furthermore, when the NNN hopping strength is not zero, this model cannot be separated into two 1D models. Numerical evidence demonstrates that it still exhibits critical phases, and we analyze the influence of the NNN hopping on the phase boundaries. Finally, we provide experimental schemes to realize this model.

\section{Model}
On a 2D square lattice, we investigate the Hamiltonian
\begin{equation}\label{1.1}
	\begin{split}
		H =&\sum_{ij}[h^{x}_{i,j}c_{i,j}^{\dagger}c_{i+1,j}+h^{y}_{i,j}c_{i,j}^{\dagger}c_{i,j+1}+h.c. +V_{i,j}c_{i,j}^{\dagger}c_{i,j}] \\
		&+\lambda_3\sum_{ij}(c_{i+1,j+1}^{\dagger}c_{i,j}+c_{i-1,j+1}^{\dagger}c_{i,j}+h.c.),
	\end{split}
\end{equation}
with
\begin{equation}\label{1.2}
	\begin{split}
		h^{x}_{i,j} =& \cos[2\pi\alpha_1(i+\frac{1}{2})+\delta_{1}]+\lambda_1, \\
		h^y_{i,j} =& \cos[2\pi\alpha_2(j+\frac{1}{2})+\delta_{2}]+\lambda_1, \\
		V_{i,j} =& \lambda_2[\cos(2\pi\alpha_1 i+\delta_{1})+\cos(2\pi\alpha_2 j+\delta_{2})],
	\end{split}
\end{equation}
where $c_{i,j}^{\dagger}(c_{i,j})$ creates (annihilates) an electron at position $(i,j)$, $\alpha_{1}$ and $\alpha_{2}$ are irrational numbers, $\delta_{1}$ and $\delta_{2}$ are arbitrary phase shift, $\lambda_1$ adjusts the nearest-neighbor hopping strength,  $\lambda_2$ is the strength of the on-site quasiperiodic potential, and $\lambda_3$ is the strength of the NNN hopping. For convenience, we fix $\alpha_1=\frac{\sqrt{5}-1}{2}$, $\alpha_2=\frac{\sqrt{2}}{2}$, $\delta_{1}=\delta_{2}=0$ and unless otherwise stated, set that the sizes in two directions are equal, i.e., $L_x=L_y=L$.

\section{Without NNN hopping}

\subsection{Phase diagram}
\begin{figure}[htbp]
	\centering
	\includegraphics[width=1\linewidth]{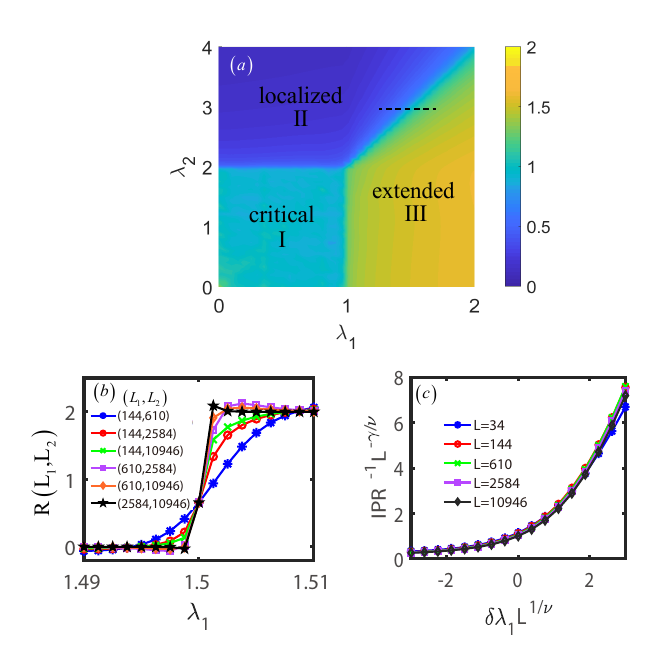}\\
	\caption{\label{01} (a) Phase diagram of the 2D model (\ref{1.1}) with $\lambda_3=0$ characterized by the fractal dimension. Here we take the system size $L=80$. The regions I, II and III correspond to the critical, localized, and extanded phases, respectively. (b)(c) The finite size scaling analysis near the localized-extended transition with $\lambda_2=3$ being fixed. (b) The plot of $R(L_1,L_2)$ as a function of $\lambda_1$ for several pairs of $(L_1,L_2)$. (c) The plot $IPR^{-1}L^{-\frac{\gamma}{\nu}}$ versus $\delta\lambda_1L^{\frac{1}{\nu}}$ for different sizes, all the curves convergence for $\nu=1$ and $\gamma=0.659\pm 0.004$.}
\end{figure}

We first discuss the phase diagram of this model when the NNN hopping strength is zero (i.e., $\lambda_3=0$).
We aim to find the solutions of eigenfunction $H|\Psi\rangle=E|\Psi\rangle$ from separation of variables. 
The Hamiltonian $H$ with $\lambda_3=0$ can be further written as $H = H_x + H_y=\sum_jH_{x,j}+\sum_iH_{y,i}$, with $H_{x,j}=\sum_{i}h^{x}_{i,j}c_{i,j}^{\dagger}c_{i+1,j}+h.c.+\lambda_{2}\sum_{i}\cos(2\pi\alpha_{1}i+\delta_1)c_{i,j}^{\dagger}c_{i,j}$ and $H_{y,i}=\sum_{j}h^{y}_{i,j}c_{i,j}^{\dagger}c_{i,j+1}+h.c.+\lambda_{2}\sum_{j}\cos(2\pi\alpha_{2}j+\delta_2)c_{i,j}^{\dagger}c_{i,j}$. 
$H_{x,j}$ with fixed $j$ and $H_{y,i}$ with fixed $i$ are two 1D extended Aubry-Andr\'{e} (EAA) models with quasiperiodic hopping and quasiperiodic on-site potentials. Hence, the 2D model (\ref{1.1}) is decoupled into two 1D EAA models ~\cite{Kohmoto1990,WangMBC,Takada,Liu}. We set $H_x|\psi_x\rangle=E_x|\psi_x\rangle$ and $H_y|\phi_y\rangle=E_y|\phi_y\rangle$ with $E_x+E_y=E$.  We see  $[H_x, H_y]=0$, then the wavefunction can be written as the tensor product
$|\Psi\rangle=|\psi_x\rangle\otimes|\phi_y\rangle$. Therefore, $|\Psi\rangle$ and $|\psi_x\rangle\otimes|\phi_y\rangle$ have similar extended, localized, and critical properties. According to the phase diagram of the EAA model, we similarly obtain the phase diagram of Hamiltonian (\ref{1.1}), as shown in Fig. \ref{01}(a), where the critical, localized, and extanded phases correspond to the region I, II and III, respectively. To show the phase diagram, we calculate the average fractal dimension, which for any arbitrary $m$-th eigenstate  $|\Psi_{m}\rangle=\sum_j^{L^2}\psi_{m,j}c^{\dagger}_j|\varnothing\rangle$ is related to the mean inverse participation ratio (MIPR) by $D=-\frac{\ln MIPR}{\ln L}$, with $MIPR=\frac{1}{L^2}\sum_{m=1}^{L^2}\sum_{j=1}^{L^2}|\psi_{m,j}|^4$. It is known that in the thermodynamic limit $L\rightarrow\infty$, the fractal dimension $D=2$ in the extended phase, $D=0$ in the localized phase, and $0<D<2$ in the critical phase [Fig. \ref{01}(a)]. The three phase boundaries are $\lambda_1=1$, $\lambda_2=2$ and $\lambda_2=2\lambda_1$, respectively.

Anderson transition is the continuous phase transition, where the scaling behavior occurs near the phase boundary, and critical exponent can be defined to characterize different types of universality classes. Near the critical point, we introduce the  critical exponents $\nu$ and $\gamma$, which describe the divergence of the correlation and localization lengths $\xi$ and IPR close to the transition as 
\begin{equation}\label{xi}
	\xi\sim |\delta\lambda_{\alpha}|^{-\nu}, \quad  IPR\sim (\delta\lambda_{\alpha})^{\gamma},
\end{equation}
 where $\delta\lambda_{\alpha}=(\lambda_{\alpha}-\lambda_{\alpha,c})/\lambda_{\alpha,c}$ with $\alpha=1,2$, and $\lambda_{\alpha,c}$ being the phase transition point.
This system can be analogized to the thermodynamic properties of the Ising model, with $M$ as the instantaneous magnetization of a system with $N$ spins. When the temperature exceeds the critical temperature, i.e., $T>T_c$, the thermal average of $M^2$ is written as $\langle M^2\rangle\propto \chi/N$, where $\chi$ is the magnetic susceptibility~\cite{QPT1,QPT2}. We know that near the transition point, $M(T)\sim |T-T_c|^{\beta}$, $\chi\sim |T-T_c|^{-\gamma}$, and the correlation length $\xi\sim |T-T_c|^{-\nu}$~\cite{QPT1,QPT2}. We can correspond the latter two to the $IPR^{-1}$ and $\xi$ introduced earlier in Eq.~(\ref{xi}), and $\langle M^2\rangle\propto \chi/N$ to 
\begin{equation}\label{xi2}
	IPR^{-1}/L^{d}\sim (-\delta\lambda_{\alpha})^{2\beta}.
\end{equation}	
From Eq.~(\ref{xi}) and Eq.~(\ref{xi2}), we can obtain $IPR\sim \xi^{-\gamma/\nu}$ and $IPR^{-1}/L^{d}\sim \xi^{-\beta/2\nu}$. 
At the transition point, where $\xi\sim L$, we have  $IPR\sim L^{-\gamma/\nu}$ and $IPR^{-1}/L^{d}\sim L^{-\beta/2\nu}$. 
Then we can directly obtain the hyperscaling law $2\beta/\nu+\gamma/\nu=d$, which is also the same as the hyperscaling law at the phase transition point of the Ising model. By analogy with the scaling relationship for magnetization in a finite-size system, $M(L,T)\sim L^{-\beta/\nu}\times g_M((T-T_c)L^{1/\nu})$, where $g_M$ is the scaling function, we assume the following finite size scaling relationship when the system is finite:
\begin{equation}\label{scaling0}
		IPR^{-1}/L^d=L^{-2\beta/\nu}F(L^{\frac{1}{\nu}}\delta\lambda_{\alpha}),
\end{equation}
where $F(x)$ is the scaling function. Using the hyperscaling law mentioned above, Eq.~(\ref{scaling0}) simplifies to:
\begin{equation}\label{scaling}
	IPR^{-1}L^{-\frac{\gamma}{\nu}}=F(L^{\frac{1}{\nu}}\delta\lambda_{\alpha}),
\end{equation}
The analogy between the scaling analysis near the phase transition point of the model we discussed and that of the Ising model is summarized in Table~\ref{wavefunction}.
\begin{table}[h]\renewcommand{\arraystretch}{1.5}
	\centering
 \begin{tabular}{|c|c|ll}
		\cline{1-2}
		Ising model  &  The model we studied  &\\ \cline{1-2}
		$\xi\sim |T-T_c|^{-\nu}$ & $\xi\sim |\delta\lambda_{\alpha}|^{-\nu}$ &\\ \cline{1-2}
		 $\chi\sim |T-T_c|^{-\gamma}$  & $IPR^{-1}\sim (\delta\lambda_{\alpha})^{-\gamma}$ &\\ \cline{1-2}
		$\langle M^2\rangle\propto \chi/N\sim |T-T_c|^{2\beta}$ & $IPR^{-1}/L^{d}\sim (-\delta\lambda_{\alpha})^{2\beta}$ & \\ \cline{1-2}
		$M\sim L^{-\beta/\nu}g_M((T-T_c)L^{1/\nu})$  & $IPR^{-1}L^{-\frac{\gamma}{\nu}}=F(L^{\frac{1}{\nu}}\delta\lambda_{\alpha})$ &\\ \cline{1-2}
	\end{tabular}
	\caption{The analogy between the scaling analysis near the phase transition point of the Ising model and that of the model we discussed.}\label{wavefunction}
\end{table}

At the transition point, we have $\delta\lambda_{\alpha}=0$, and thus Eq.~(\ref{scaling}) becomes $IPR^{-1}=L^{\gamma/\nu}F(0)$. Then a function of two size-variables can be defined as~\cite{FSZ1,FSZ2}:
\begin{equation}\label{scaling2}
	R[L_1,L_2]=\frac{\ln(IPR_2/IPR_1)}{\ln(L_1/L_2)},
\end{equation}
which equals to $\gamma/\nu$ at the critical point for any pair $(L_1, L_2)$, where $L_1$ and $L_2$ represent two different system sizes. Fig.~\ref{01}(b) displays the behaviors of $R[L_1, L_2]$ as a function $\lambda_1$ for different pairs of $L_1$ and $L_2$ with fixed $\lambda_2=3$. From the crossing point, we can determine the critical point $\lambda_{1c}=1.500\pm 0.006$, which is consistent with phase boundary $\lambda_{2}=2\lambda_{1}$, and the corresponding critical exponent $\gamma/\nu\approx 0.66$.  In Ref.~\cite{WangZhang}, we analytically obtained the localization length of a two-dimensional quasiperiodic vertex-decorated Lieb lattice and found the critical exponent $\nu = 1$. Therefore, we directly choose $\nu = 1$ here and plot $IPR^{-1}L^{-\frac{\gamma}{\nu}}$ as a function of $\delta\lambda_{1}L^{\frac{1}{\nu}}$ for different size in Fig. \ref{01}(c), we see all the curves superposed together, and the critical exponent $\gamma=0.659\pm 0.004$~\cite{errorbar},
which is approximately twice the critical exponent of the Anderson transition induced by quasiperiodic potential in one dimension (According to the definition in Eq.~(\ref{xi}), one can obtain the critical exponent $\gamma\approx 0.33$
near the Anderson transition point in a 1D quasiperiodic system~\cite{HYao2019}). Using the same approach, we can determine the transition points and critical exponents for the transition from critical phase to localized or extended phase.

%Fisher propose a scaling relationship:

\subsection{Dynamics of Wave Packet and Transport Properties}
\begin{figure}[htbp]
	\centering
	\includegraphics[width=1\linewidth]{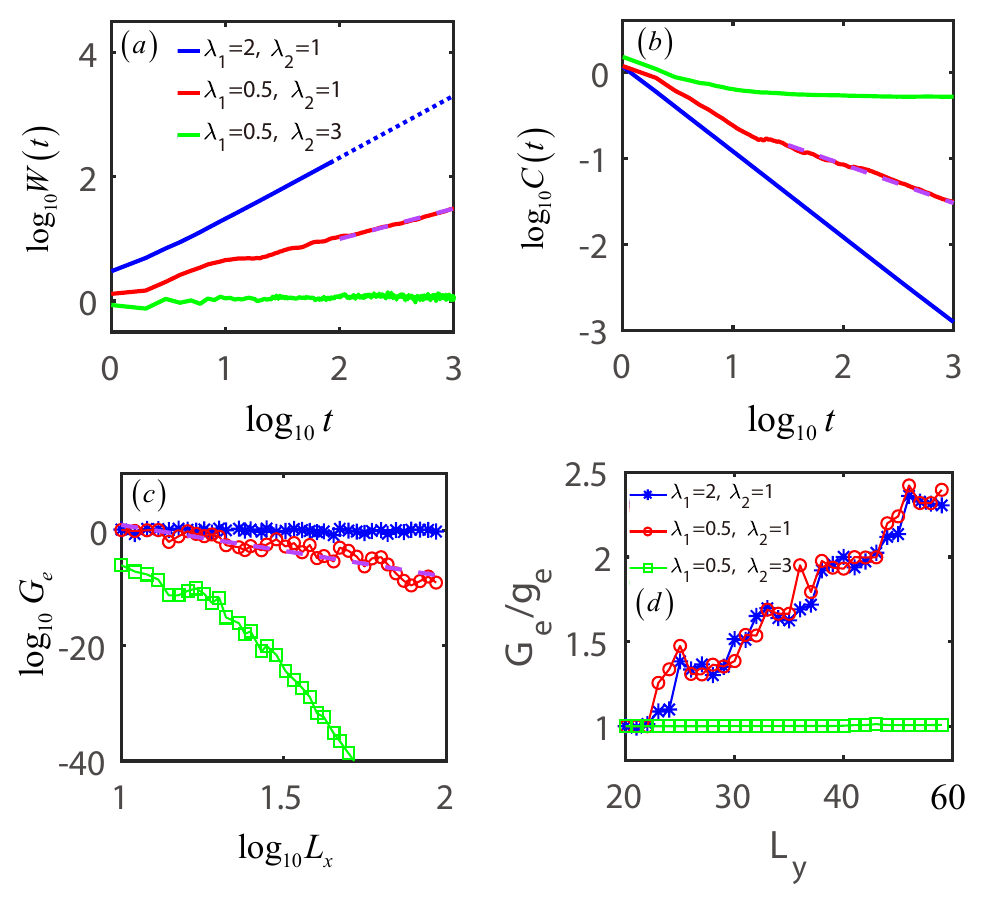}\\
	\caption{The evolution of (a) mean square displacement and (b) autocorrelation function averaged over $50$ samples. The particle is initially localized at $(i,j)=(\frac{L_x}{2},\frac{L_y}{2})$ with system size $L_x=L_y=400$. (c) The conductance as a function of $L_x$ in the unit of $e^2/h$ with $L_y=10$ fixed. (d)The conductance as a function of $L_y$ in the unit of $g_{e}=G_{e}(L_y=20)$ with $L_x=10$ fixed. The blue/red/green curves in (a)-(d) are in the extended, critical and localized phases, with the same values to the legend in (a). The purple dashed line in (a)-(c) is the linear fitting in the critical phase. Here $\lambda_3=0$.}\label{02}
\end{figure}

We now discuss the diffusion dynamics of the wave packet, which can be measured directly from the experiments. Suppose a particle is initially localized at site $(p,q)$ and evolves according to the Hamiltonian. We study the mean square displacement
\begin{equation}\label{3.1}
	W(t)=\sqrt{\sum_{i,j}\bigg[(p-i)^2+(q-j)^2\bigg]\langle n_{i,j}(t)\rangle}
\end{equation}
and the smoothed autocorrelation function
\begin{equation}\label{3.2}
	C(t)=\frac{1}{t}\int_{0}^{t}|\langle\Psi(0)|\Psi(t^{'})\rangle|^2dt^{'},
\end{equation}
where $\langle n_{i,j}(t)\rangle$ is the time dependent particle numbers at site $(i,j)$ and $|\Psi(t)\rangle$ is the wavefunction at time $t$. Then the long time dynamical behavior of $W(t)$ and $C(t)$ are
\begin{equation}\label{3.3}
%	\begin{split}
		W(t) \sim t^{\kappa}, \quad 
		C(t) \sim t^{-\beta},
%	\end{split}
\end{equation}
where the $\kappa$ and $\beta$ depend on the energy spectrum and fractal characteristics of the wavefunctions \cite{Hiramoto,ZhongJX,Ketzmerick}. Figs. \ref{02}(a) and (b) show the evolution of $W(t)$ and $C(t)$. The parameters of blue, red, and green lie in the extended, critical, and localized phases, respectively. The dynamical index $\kappa$  is similar to the 1D case: $\kappa = 1$ signifies the ballistic evolution in the extended phase, $\kappa = 0$ indicates localization in the localized phase, and $\kappa\approx 1/2$ suggests approximately normal diffusive behavior at the critical point. Then we consider the magnitude of $\beta$. In localized phases, the exponent $\beta$ behaves similarly to the 1D case, being $0$. However, in critical and extended phases, $\beta$ shows dimension dependence. In the extended phase, previous studies on the 1D case suggested $\beta$ to be approximately $0.84$~\cite{Ketzmerick}, or the relationship between $C(t)$ and $t$ could be $C(t)\sim\ln(t)/t$ \cite{JXZhong}.
However, in the 2D case, the relationship between $C(t)$ and $t$ satisfies $C(t)\sim 1/t$, meaning $\beta=1$. At the critical phase or critical point, $\beta\approx 0.14\sim 0.25$ in 1D systems, but in our study of the 2D model, $\beta$ increased to about $0.45$. $C(t)$ can be considered as measuring the probability that the system remains in its initial state, and 2D systems have more channels to escape the initial state than 1D systems. Therefore, in the extended or critical phases, $\beta$ in two dimensions is greater than in 1D systems.

% The value of $\kappa$ and $\beta$ is one in the extended phase and zero in the localized phase, and we note that different from the one dimensional case, the autocorrelation do not has a logarithmic contribution as predicted in \cite{Zhong1}. In the extended phase, the wavefunction will reach the boundary quickly as the size of system is finite, hence we use the dotted line the show the dynamics in the infinite size limit. In the critical phase, numerical fitting shows that $\kappa\sim0.36$ and $\beta\sim0.47$ as plotted in the purple dashed line.

%\section{Transport Properties}
Then we investigate the transport properties for the 2D model by contacting the sample with two leads from the left and right side in the x-direction. In the zero temperature limit and linear response regime, the conductance can be written as~\cite{Mahan}
\begin{equation}\label{4.1}
	G_{e}=\frac{e^2}{h}Tr[\mathbf{\Gamma}_{L}\mathbf{G}^{A}\mathbf{\Gamma}_{R}\mathbf{G}^{R}],
\end{equation}
where the Green's function $\mathbf{G}^{R}(\omega)=[\omega-\mathbf{H}-\mathbf{\Sigma}_{L}^{R}(\omega)-\mathbf{\Sigma}_{R}^{R}(\omega)]^{-1}$ and $\mathbf{G}^{A}(\omega)=[\mathbf{G}^{R}(\omega)]^{\dagger}$, and the spectral density $\mathbf{\Gamma}_{L(R)}=i(\mathbf{\Sigma}_{L(R)}^{R}-\mathbf{\Sigma}_{L(R)}^{A})$. The self energies induced by the leads are $\Sigma_{L,j1}(\omega)=\Sigma_{R,jL}(\omega)=-i\frac{\gamma}{2}$, which are energy independent and non-vanishing only at the leftmost or the rightmost column in the wide band limit.

Fig. \ref{02}(c) shows the conductance as a function of the length $L_x$ with $L_y=10$, and Fig. \ref{02}(d) displays the conductance as a function of the width $L_y$ with $L_x=10$. The blue, red and green curves are selected in the extended, critical and localized phases, with the parameters the same as Figs. \ref{02}(a)(b). In the extended phase, the transport is ballistic in the bulk, and hence the conductance is independent of length $G_{e}\sim L_x^0$ and proportional to the numbers of scattering channels and hence $G_{e}\sim L_y^1$. In the critical phase, the transport is diffusive indicating the power-law dependence of length $G_{e}\sim L_x^{-\alpha}$ and interestingly, also proportional to the width $G_{e}\sim L_y^{1}$ just like the extended phase. In the localized phase, the bulk is insulating and the conductance is exponentially decay with respect to the length $G_{e}\sim e^{-L_x}$ and independent of the width $G_{e}\sim L_y^{0}$. In all the cases, the chemical potentials (Fermi energies) are selected inside the energy band. 
%In Fig. \ref{Fig2}(c) the conductance is in the unit of $\frac{e^2}{h}$, and in Fig. \ref{Fig2}(d) the conductance is in the unit of $g_{e}=G_{e}(L_y=20)$.

\begin{figure}[htbp]
	\centering
	\includegraphics[width=1\linewidth]{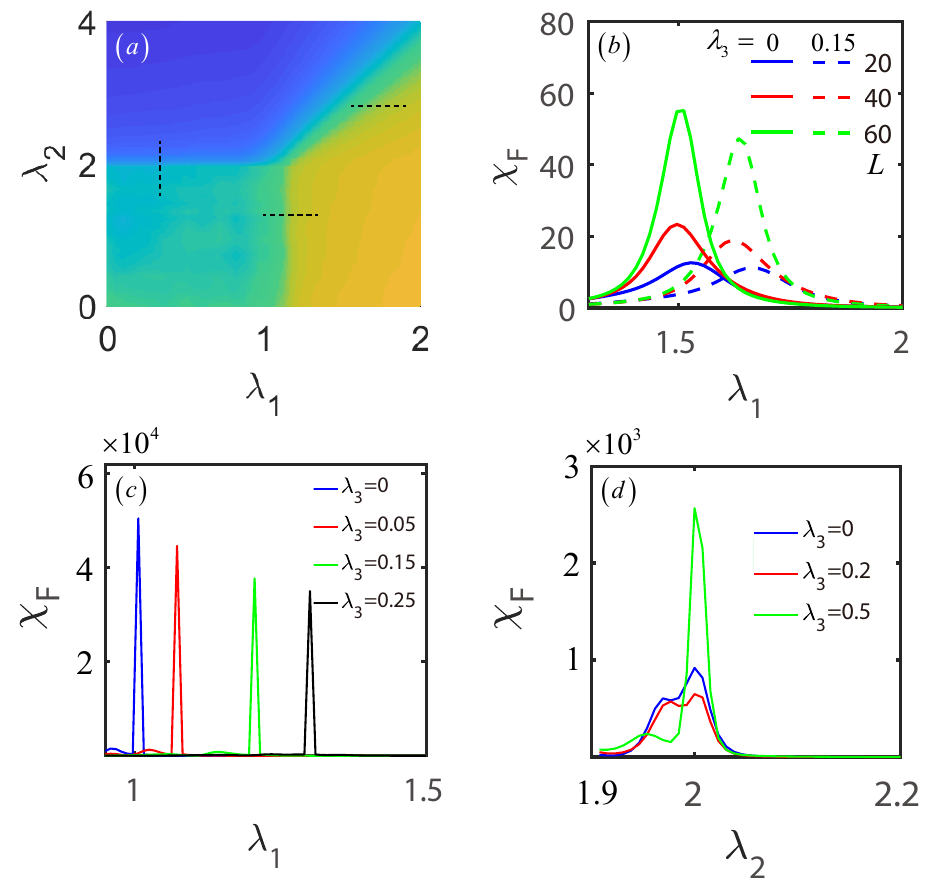}\\
	\caption{(a) Phase diagram of the 2D model characterized by the average fractal dimension with NNN hopping $\lambda_3=0.15$, and the colourmap is the same as Fig. \ref{01}(a). (b) Fidelity susceptibility as a function of $\lambda_1$ for different sizes with fixed $\lambda_2=3$. The NNN hopping $\lambda_3=0$ for solid lines and $\lambda_3=0.15$ for dashed lines. (c) Fidelity susceptibility versus $\lambda_1$ with $\lambda_2=1.25$. (d)Fidelity susceptibility versus $\lambda_2$ for $\lambda_1=0.3$. The system sizes here are $L_x=L_y=40$.}\label{Fig3}
\end{figure}

\section{With NNN hopping}
In this section, we discuss whether the critical phase is stable when the NNN hopping strength $\lambda_3$ is non-zero. The 2D model (\ref{1.1}) with non-zero $\lambda_3$ cannot be separated into two 1D models.
Fig. \ref{Fig3}(a) shows the phase diagram using the average fractal dimension with $\lambda_3=0.15$. It can be seen there are still the extended, critical and localized phases, but the phase boundary will change. 

To see how the phase boundary shift, it is convenient to calculate the fidelity susceptibility
\begin{equation}\label{5.2}
	\chi_{F}=-2\sum_{\alpha=1,2}\lim_{\delta\lambda_{\alpha}\rightarrow0}\frac{\ln F_{\alpha}}{\delta\lambda_{\alpha}^2},
\end{equation}
where the ground state fidelity $F_{\alpha}=|\langle\Psi_{g}(\lambda_{\alpha}+d\lambda_{\alpha})|\Psi_{g}(\lambda_{\alpha})\rangle|$ measures the overlap between $|\Psi_{g}(\lambda_{\alpha})\rangle$ and $|\Psi_{g}(\lambda_{\alpha}+d\lambda_{\alpha})\rangle$ \cite{Zanardi,GuSJ,ChenS}. Near the phase transition point, the behavior of wavefunctions changes dramatically and $\chi_{F}$ will be divergent. Fig. \ref{Fig3}(b) shows the $\chi_F$ as a function of $\lambda_1$ for different sizes with $\lambda_2=3$. It can be seen that without the NNN hopping $\lambda_3=0$, $\chi_F$ has a peak at $\lambda_1=1.5$, which gives the same phase boundary as the fractal dimension. In the presence of $\lambda_3$, the peak shifts to $\lambda_1\sim 1.63$, and in both cases, the peak will be divergent as increasing the system size. Figs. \ref{Fig3}(c) and (d) show the fidelity susceptibility near the transition points from the critical phase to extended and localized phases, respectively, with different $\lambda_3$.
As the increase of $\lambda_3$, the peak of $\chi_F$ shifts towards large $\lambda_1$ [Fig. \ref{Fig3}(c)], while the peak of $\chi_F$ gives nearly the same $\lambda_2$ [Fig. \ref{Fig3}(d)]. Therefore, the phase boundary between critical and extended phases and the boundary between localized and extended phases shift toward larger $\lambda_1$, while the phase boundary between critical and localized phases remains $\lambda_2=2$, as shown in Fig. \ref{Fig3}(a).
 
\

\section{Experimental realization}
\begin{figure}[htbp]
 	\centering
 	\includegraphics[width=0.65\linewidth]{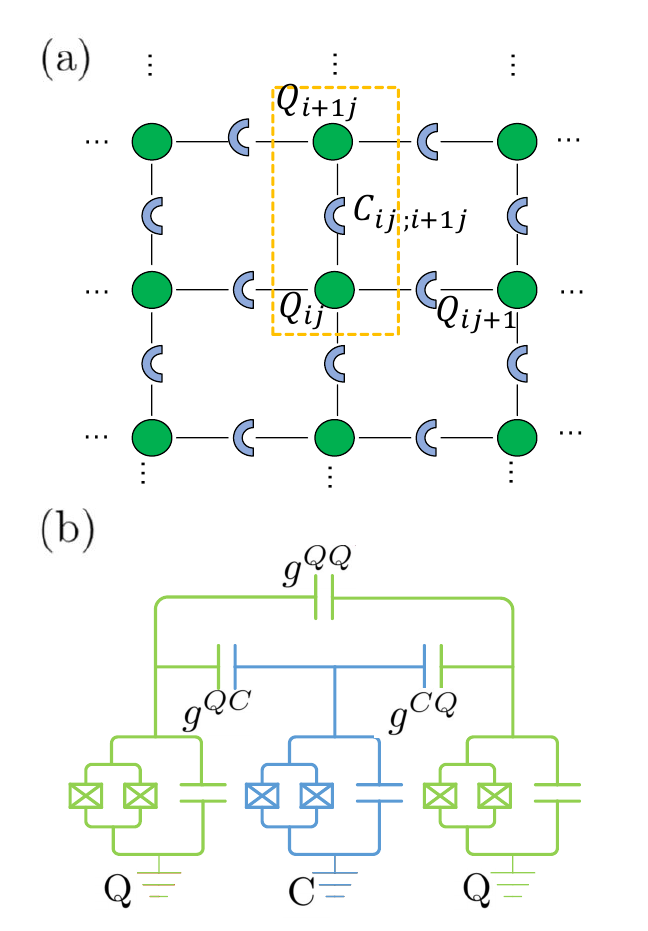}\\
 	\caption{(a) Schematic representation of the 2D model (\ref{1.1}) with $\lambda_3=0$ on the superconducting quantum processor, which consists of $L_x\times L_y$ transmon superconducting qubits and $(L_x-1)L_y+(L_y-1)L_x$ tunable couplers. (b) The circuit diagram of neighboring qubits and the coupler between them.}\label{Fig4}
\end{figure}

Finally, we discuss the implementation of the 2D model (\ref{1.1}), whose hopping strength is quasiperiodic, which is difficult to achieve in some simulated systems. Here, we discuss the implementation of this model on a
a superconducting quantum processor (SQP), using the special case of $\lambda_3=0$ as an example. The case with $\lambda_3\neq0$ can be realized in a similar manner. The SQP consists of the 2D array of $L_x\times L_y$ transmon superconducting qubits, along with $(L_x-1)L_y+(L_y-1)L_x$ tunable couplers, each positioned between every two nearest-neighbor qubits, as shown in Fig. \ref{Fig4}(a). The effective coupling coupling $h^{x}_{i,j}$ in Eq. (\ref{1.1}) can be described by~\cite{YanF,HengFan2021}:
\begin{equation}
h^x_{i,j}=g^{QQ}_{ij;i+1j}+\frac{g^{QC}_{ij;iji+1j}g^{QC}_{i+1j;iji+1j}}{\Delta_{ij,i+1j}},
\end{equation}
%For the sake of convenience in notation, here we uniformly omit the index $j$ for the $y$-axis. The coupling
which contains two parts: the first term $g^{QQ}_{ij;i+1j}$ represents the direct coupling between neighboring qubits $Q_{i,j}$ and $Q_{i+1,j}$, and the second term represents the coupling between the coupler $C_{ij,i+1j}$ and the two nearest-neighbor qubits $Q_{i,j}$ and $Q_{i+1,j}$ (see Fig. \ref{Fig4}), where $g^{QC}_{ij;iji+1j}$ is the coupling strength between $Q_{i,j}$ and $C_{ij,i+1j}$, and $1/\Delta_{ij,i+1j}=[1/(\omega_{ij}-\omega^{C}_{ij;i+1j})+1/(\omega_{i+1j}-\omega^{C}_{ij;i+1j})]/2$ with $\omega_{ij}$ and $\omega^{C}_{ij;i+1j}$ being the corresponding frequency of the qubit $Q_{i,j}$ and the coupler $C_{ij,i+1j}$, respectively.  Similarly, we can obtain the hopping strength $h^y_{i,j}$ in the $y$-direction. The on-site quasiperiodic potential $V_{i,j}$ is represented as:
\begin{equation}
	\begin{split}
	V_{i,j}=&\omega_{ij}+\frac{(g_{ij;iji+1j}^{QC})^{2}}{\omega_{ij}-\omega_{ij;i+1j}^{c}}+
	\frac{(g_{ij;iji-1j}^{QC})^{2}}{\omega_{ij}-\omega_{ij;i-1j}^{c}} \\
	+&\frac{(g_{ij;ijij+1}^{QC})^{2}}{\omega_{ij}-\omega_{ij;ij+1}^{c}}+
	\frac{(g_{ij;ijij-1}^{QC})^{2}}{\omega_{ij}-\omega_{ij;ij-1}^{c}}.
    \end{split}
\end{equation}
By tuning the frequencies of qubits and couplers, we can obtain the desired forms and strengths of the nearest-neighbor hopping and on-site potential.
 
By manipulating both qubits and couplers, we can experimentally observe the dynamics of different phases in this model. Therefore, we can experimentally detect the critical phase and transitions between different phases based on their dynamic properties.

\section{Conclusion and Discussion} 
We have introduced a 2D quasiperiodic model exhibiting a critical phase and provided its phase diagram.
We adapted a finite-size scaling method to effectively determine the phase boundaries of this model and obtain critical exponents. Additionally, we explored the properties of the critical, extended and localized phases in terms of wave packet diffusion and transport in this 2D system, and examined the changes of phase boundaries in the phase diagram induced by the introduction of NNN hopping through calculations of fidelity susceptibility. Finally, we discussed the realization of this model in superconducting circuits. Our work paves the way for searching the critical phase in 2D systems.

Recent research has revealed the presence of novel anomalous MEs in 1D systems. Unlike conventional MEs that separate extended states from localized states, these MEs delineate the critical state from either extended or localized states~\cite{XDeng2019,Wang2022,XCZhou,LiuT}. Our findings suggest that since critical phases can be extended from one dimension to two dimension, these novel MEs can also be extended to two dimension, warranting further investigation in the future.

%\section*{Conflict of interest}
%The authors declare that they have no conflict of interest.

\begin{acknowledgements}
This work is supported by National Key R\&D Program of China under Grant No.2022YFA1405800, the National Natural Science
Foundation of China (Grant No.12104205), the Key-Area Research and Development Program of Guangdong Province (Grant No. 2018B030326001), Guangdong Provincial Key Laboratory (Grant No.2019B121203002).
\end{acknowledgements}
			
%%%%%%%%%%%%%%%%%%%%%%%%%%%%%%%%%%%%%%%%%%%%%

%%%%%%%%%%%%%%%%%%%%%%%%%%%%%%%%%%%%%%%%%%%%%%%%%%%%%

\begin{thebibliography}{36}
\bibitem{noteadd} We note that we are considering a system where the potential is quasiperiodic, but the spatial structure is periodic.
\end{thebibliography}
\end{document}